\documentclass[twocolumn,letter,epsf]{jpsj2}
\usepackage[dvips]{color}
\usepackage{bm}
\usepackage{pstricks}%
\usepackage{pst-eps}%
\usepackage{pst-plot}%
\usepackage{pst-node}%

\renewcommand{\i}{{\rm i}}
\newcommand{\e}{{\rm e}}

\title{Exact Plaquette-Ordered Ground States in the Generalized Hubbard
Model\\ in Arbitrary Dimensions}
\author{Masaaki {\sc Nakamura}$^1$ and Kazuhito {\sc Itoh}$^2$}
\inst{
$^1$Department of Applied Physics, Faculty of Science, Tokyo
University of Science, Tokyo 162-8601\\
$^2$Department of Complex Systems Science, Graduate School of
Information Science, Nagoya University, Nagoya 464-8601}
\kword{Hubbard model, site-off-diagonal interactions, exact ground
 states, bond-ordered state, plaquette-ordered state, Kagom\'e lattice,
 Pyrochlore lattice
}

\recdate{}
\abst{
 We show the existance of the exact plaquette-ordered ground states of
  the Hubbard model including site-off-diagonal interactions in
  arbitrary dimensions, by decomposing the Hamiltonian as sum of
  products of projection operators for each spin sector.  The obtatined
  exact ground states are interpreted as N\'eel ordered states on the
  dual lattices. We demonstrate this idea in the one-dimensional chain
  and higher-dimensional lattices such as the Kagom\'e lattice, and
  determine parameter regions of the exact ground states.
}
\begin{document}
\sloppy \maketitle

\section{Introduction}

The Hubbard model is one of the generic models to describe interacting
electrons in narrow-band systems~\cite{Hubbard}.  This model has played
central roles to study magnetism and superconductivity.  In spite of its
simplicity, it is difficult to solve this model exactly except for one
dimension or some special cases.  On the other hand, it has been
considered many extensions of the Hubbard model.  The on-site repulsion
of this model is due to the matrix elements of the Coulomb interaction
corresponding to the on-site Wannier states, and other matrix elements
are neglected. Therefore, we consider effects of these neglected terms
as site-off-diagonal interactions.\cite{Campbell-G-L} For these
generalized models, exact results for ferromagnetism and superconducting
states have been
discussed.\cite{Strack-V1993,Strack-V1994,Arrachea-A,Boer-K-S,
Boer-S,Montorsi-C,Kollar-S-V}

Few years ago the authors discussed a different type of exact ground
state in a one-dimensional system that is ``bond N\'eel''
state,\cite{Itoh-N-M} using the decomposition of the Hamiltonian into
projection operators \cite{Itoh}. Furthermore they suggested extensions
of this argument to higher dimensional systems.\cite{Itoh-N-M} Main aim
of this paper is to demonstrate the existence of the exact ground states
in the generalized Hubbard model in arbitrary dimensions, especially
that on the Kagom\'e lattice where few results have been known for
Hubbard-type models.\cite{Tanaka-U,Imai-K-T}

The rest of this paper is organized as follows: In
sec.~\ref{sec:method}, we review the method to construct Hamiltonians
with exact ground states.  In sec.~\ref{sec:1D}, we apply this method to
the one-dimensional model discussed in ref.~\ref{Itoh-N-M}, and obtain
the extended version of the phase diagram.  In sec.~\ref{sec:2D}, we
apply the analysis to the Kagom\'e lattice.  Finally, we give summary
and discussion of the results.

\section{Method}\label{sec:method}

 The strategy to construct a Hamiltonian with an exact ground state is
 the following way.\cite{Itoh} First, we consider a Hamiltonian that can
 be decomposed as sum of products of operators for different (spin)
 sectors as,
 \begin{equation}
  {\cal H}=\sum_{\alpha} h_{\alpha},\quad
   h_{\alpha}=\sum_{\mu,\nu}\lambda_{\mu\nu}
   R^{(\mu)}_{\alpha\uparrow}R^{(\nu)}_{\alpha\downarrow},
   \quad
   \lambda_{\mu\nu}\geq 0,
   \label{Ham}
 \end{equation}
 where $\alpha$ denotes the position of one of the unit plaquettes that
 cover the lattice.  $R^{(\mu)}_{\alpha\sigma}$
 ($\sigma=\uparrow,\downarrow$) is an operator satisfying
 $[R^{(\mu)}_{\alpha\uparrow},R^{(\nu)}_{\alpha\downarrow}]=0$. The
 expectation value of this operator is nonnegative $\langle
 R^{(\mu)}_{\alpha\sigma}\rangle\geq 0$. This condition is realized, if
 $R^{(\mu)}_{\alpha\sigma}$ is given by a product of an operator and its
 Hermitian conjugate. Then the expectation value of the Hamiltonian is
 also nonnegative $\langle {\cal H}\rangle\geq 0$.

 Next, we introduce a trial wave function given by a direct product of
 up and down spin sectors,
 \begin{equation}
  |\Psi({\cal A},{\cal B})\rangle=|\Phi_{\uparrow}({\cal A})\rangle
   \otimes|\Phi_{\downarrow}({\cal B})\rangle,\label{state}
 \end{equation}
 where ${\cal A}$ and ${\cal B}$ denote two groups of plaquettes that
 cover the lattice satisfying ${\cal A}\cup{\cal B}=\{\mbox{all lattice
 sites}\}$.  We require that the projection operators have the following
 conditions,
 \begin{equation}
  R^{(\mu)}_{\alpha\uparrow}|\Phi_{\uparrow}({\cal A})\rangle
   =R^{(\mu)}_{\beta\downarrow}|\Phi_{\downarrow}({\cal B})\rangle=0,
 \end{equation}
 where $\alpha\in{\cal A}$ and $\beta\in{\cal B}$. Note that
 \begin{equation}
  R^{(\mu)}_{\beta\uparrow}
   |\Phi_{\uparrow}({\cal A})\rangle
   \neq 0,\quad
   R^{(\mu)}_{\alpha\downarrow}
   |\Phi_{\downarrow}({\cal B})\rangle\neq 0.
 \end{equation}
 Therefore, the eigenvalue of the Hamiltonian for $|\Psi({\cal A},{\cal
 B})\rangle$ is always zero. Then, the lower bound and the upper bound
 of the energy coincide, so that $|\Psi({\cal A},{\cal B})\rangle$ turns
 out to be one of the exact ground state of this system.

\begin{figure}[t]
 \begin{center}
  \begin{pspicture}(0,0)(4,1.0)
   \psset{unit=8mm}
   \multirput(0,0)(2,0){2}{
   \rput(0,0.3){
   \psline[linestyle=dashed,linewidth=1.2pt](0,0)(1,0)
   \psline[linestyle=solid,linewidth=1.2pt](1,0)(2,0)
   }}
   \multirput(0,0)(1,0){5}{
   \rput(0,0.3){
   \psdots[dotstyle=o](0,0)
   }}
   \rput[c](1,0){$i$}
   \rput[c](2,0){$j$}
   \rput[c](0.5,0.6){$\downarrow$}
   \rput[c](1.5,0.6){$\uparrow$}
   \rput[c](2.5,0.6){$\downarrow$}
   \rput[c](3.5,0.6){$\uparrow$}
  \end{pspicture}
  \mbox{Chain}

  \begin{displaymath}
   \begin{array}{cc}
    \begin{pspicture}(0,-1.0)(3,3.5)
     \psset{unit=8mm}
     \multirput(0,0)(2,0){2}{
     \rput(0,0){
     \pspolygon[linestyle=dashed,linewidth=1.2pt]
     (0,0)(1,0)(0.5,0.8660254)
     \psline[linewidth=0.3pt]
     (0.5,0.28867513)(0,0)
     (0.5,0.28867513)(0.5,0.8660254)
     (0.5,0.28867513)(1,0)
     }
     \rput(0,1.7320508){
     \pspolygon[linestyle=solid,linewidth=1.2pt]
     (0,0)(1,0)(0.5,-0.8660254)
     \psline[linewidth=0.3pt]
     (0.5,-0.28867513)(0,0)
     (0.5,-0.28867513)(0.5,-0.8660254)
     (0.5,-0.28867513)(1,0)
     }
     \rput(1,1.7320508){
     \pspolygon[linestyle=dashed,linewidth=1.2pt]
     (0,0)(1,0)(0.5,0.8660254)
     \psline[linewidth=0.3pt]
     (0.5,0.28867513)(0,0)
     (0.5,0.28867513)(0.5,0.8660254)
     (0.5,0.28867513)(1,0)
     }
     \rput(1,3.4641016){
     \pspolygon[linestyle=solid,linewidth=1.2pt]
     (0,0)(1,0)(0.5,-0.8660254)
     \psline[linewidth=0.3pt]
     (0.5,-0.28867513)(0,0)
     (0.5,-0.28867513)(0.5,-0.8660254)
     (0.5,-0.28867513)(1,0)
     }
     \rput(0,3.4641016){
     \pspolygon[linestyle=dashed,linewidth=1.2pt]
     (0,0)(1,0)(0.5,0.8660254)
     \psline[linewidth=0.3pt]
     (0.5,0.28867513)(0,0)
     (0.5,0.28867513)(0.5,0.8660254)
     (0.5,0.28867513)(1,0)
     }
     \rput(1,0){
     \pspolygon[linestyle=solid,linewidth=1.2pt]
     (0,0)(1,0)(0.5,-0.8660254)
     \psline[linewidth=0.3pt]
     (0.5,-0.28867513)(0,0)
     (0.5,-0.28867513)(0.5,-0.8660254)
     (0.5,-0.28867513)(1,0)
     }
     }
     \multirput(0,0)(2,0){2}{
     \rput(0,0){
     \psdots[dotstyle=o](0,0)(1,0)(0.5,0.8660254)
     }
     \rput(0,1.7320508){
     \psdots[dotstyle=o](0,0)(1,0)(0.5,-0.8660254)
     }
     \rput(1,1.7320508){
     \psdots[dotstyle=o](0,0)(1,0)(0.5,0.8660254)
     }
     \rput(1,3.4641016){
     \psdots[dotstyle=o](0,0)(1,0)(0.5,-0.8660254)
     }
     \rput(0,3.4641016){
     \psdots[dotstyle=o](0,0)(1,0)(0.5,0.8660254)
     }
     \rput(1,0){
     \psdots[dotstyle=o](0,0)(1,0)(0.5,-0.8660254)
     }
     }
     \rput[r](2,1.5){$i$}
     \rput[l](3,1.5){$j$}
     \rput[l](2.6,0.9){$k$}
    \end{pspicture}
&
\begin{pspicture}(-0.8,-1)(4,3.5)
 \psset{unit=8mm}
 \multirput(0,0)(2,0){2}{\multirput(0,0)(0,2){2}{
 \pspolygon[linestyle=dashed,linewidth=1.2pt]
 (0,0)(0,1)(1,1)(1,0)
 \rput(1,1){
 \pspolygon[linestyle=solid,linewidth=1.2pt]
 (0,0)(0,1)(1,1)(1,0)}
 \multirput(0,0)(1,1){2}{
 \psline[linewidth=0.3pt]{-}(0,0)(1,1)
 \psline[linewidth=0.3pt]{-}(0,1)(1,0)}
 }}
 \multirput(0,0)(2,0){2}{\multirput(0,0)(0,2){2}{
 \psdots[dotstyle=o](0,0)(0,1)(1,1)(1,0)
 \rput(1,1){
 \psdots[dotstyle=o](0,0)(0,1)(1,1)(1,0)}
 }}
 \rput[r](0.95,1.8){$i$}
 \rput[l](2.05,1.8){$j$}
 \rput[l](2.05,1.2){$k$}
 \rput[r](0.95,1.2){$l$}
\end{pspicture}
\\
    \mbox{Kagom\'e}& \mbox{Checkerboard}
   \end{array}
  \end{displaymath}
	     \begin{pspicture}(2.5,-1.0)(5,3.3)
	      \psset{unit=10mm}
	      %
	      \multirput(0,2.1320508)(2,0){2}{
	      \pspolygon[linestyle=dashed,linewidth=0.6pt]
	      (2,0)(3,0)(2.5,0.8660254)
	      \psline[linestyle=dashed,linewidth=0.6pt]
	      (2.5,0.2)(2,0)
	      (2.5,0.2)(2.5,0.8660254)
	      (2.5,0.2)(3,0)
	      }
	      \multirput(1,2.1320508)(2,0){2}{
	      \pspolygon[linestyle=solid,linewidth=0.6pt]
	      (2,0)(3,0)(2.5,-0.8660254)
	      \psline[linestyle=solid,linewidth=0.6pt]
	      (2.5,-0.2)(2,0)
	      (2.5,-0.2)(2.5,-0.8660254)
	      (2.5,-0.2)(3,0)
	      }
	      \multirput(0,0.4)(2,0){2}{
	      \pspolygon[linestyle=solid,linewidth=0.6pt]
	      (2,0)(3,0)(2.5,-0.8660254)
	      \psline[linestyle=solid,linewidth=0.6pt]
	      (2.5,-0.2)(2,0)
	      (2.5,-0.2)(2.5,-0.8660254)
	      (2.5,-0.2)(3,0)
	      }
	      \multirput(0,1.7320508)(2,0){2}{
	      \pspolygon[linestyle=solid,linewidth=1.2pt]
	      (2,0)(3,0)(2.5,-0.8660254)
	      \psline[linestyle=solid,linewidth=1.2pt]
	      (2.5,-0.2)(2,0)
	      (2.5,-0.2)(2.5,-0.8660254)
	      (2.5,-0.2)(3,0)
	      }
	      \multirput(1,1.7320508)(2,0){2}{
	      \pspolygon[linestyle=dashed,linewidth=1.2pt]
	      (2,0)(3,0)(2.5,0.8660254)
	      \psline[linestyle=dashed,linewidth=0.6pt]
	      (2.5,0.2)(2,0)
	      (2.5,0.2)(2.5,0.8660254)
	      (2.5,0.2)(3,0)
	      }
	      \multirput(0,0)(2,0){2}{
	      \pspolygon[linestyle=dashed,linewidth=1.2pt]
	      (2,0)(3,0)(2.5,0.8660254)
	      \psline[linestyle=dashed,linewidth=0.6pt]
	      (2.5,0.2)(2,0)
	      (2.5,0.2)(2.5,0.8660254)
	      (2.5,0.2)(3,0)
	      }
	      \multirput(1,0.4)(2,0){2}{
	      \pspolygon[linestyle=dashed,linewidth=0.6pt]
	      (2,0)(3,0)(2.5,0.8660254)
	      \psline[linestyle=dashed,linewidth=0.6pt]
	      (2.5,0.2)(2,0)
	      (2.5,0.2)(2.5,0.8660254)
	      (2.5,0.2)(3,0)
	      }
	      \multirput(1,0)(2,0){2}{
	      \pspolygon[linestyle=solid,linewidth=1.2pt]
	      (2,0)(3,0)(2.5,-0.8660254)
	      \psline[linestyle=solid,linewidth=1.2pt]
	      (2.5,-0.2)(2,0)
	      (2.5,-0.2)(2.5,-0.8660254)
	      (2.5,-0.2)(3,0)
	      }
	      \multirput(0,2.1320508)(2,0){2}{
	      \psdots[dotstyle=o]
	      (2,0)(3,0)(2.5,0.8660254)
	      (2.5,0.2)(2,0)
	      (2.5,0.2)(2.5,0.8660254)
	      (2.5,0.2)(3,0)
	      }
	      \multirput(1,2.1320508)(2,0){2}{
	      \psdots[dotstyle=o]
	      (2,0)(3,0)(2.5,-0.8660254)
	      (2.5,-0.2)(2,0)
	      (2.5,-0.2)(2.5,-0.8660254)
	      (2.5,-0.2)(3,0)
	      }
	      \multirput(0,0.4)(2,0){2}{
	      \psdots[dotstyle=o]
	      (2,0)(3,0)(2.5,-0.8660254)
	      (2.5,-0.2)(2,0)
	      (2.5,-0.2)(2.5,-0.8660254)
	      (2.5,-0.2)(3,0)
	      }
	      \multirput(0,1.7320508)(2,0){2}{
	      \psdots[dotstyle=o]
	      (2,0)(3,0)(2.5,-0.8660254)
	      (2.5,-0.2)(2,0)
	      (2.5,-0.2)(2.5,-0.8660254)
	      (2.5,-0.2)(3,0)
	      }
	      \multirput(1,1.7320508)(2,0){2}{
	      \psdots[dotstyle=o]
	      (2,0)(3,0)(2.5,0.8660254)
	      (2.5,0.2)(2,0)
	      (2.5,0.2)(2.5,0.8660254)
	      (2.5,0.2)(3,0)
	      }
	      \multirput(0,0)(2,0){2}{
	      \psdots[dotstyle=o]
	      (2,0)(3,0)(2.5,0.8660254)
	      (2.5,0.2)(2,0)
	      (2.5,0.2)(2.5,0.8660254)
	      (2.5,0.2)(3,0)
	      }
	      \multirput(1,0.4)(2,0){2}{
	      \psdots[dotstyle=o]
	      (2,0)(3,0)(2.5,0.8660254)
	      (2.5,0.2)(2,0)
	      (2.5,0.2)(2.5,0.8660254)
	      (2.5,0.2)(3,0)
	      }
	      \multirput(1,0)(2,0){2}{
	      \psdots[dotstyle=o]
	      (2,0)(3,0)(2.5,-0.8660254)
	      (2.5,-0.2)(2,0)
	      (2.5,-0.2)(2.5,-0.8660254)
	      (2.5,-0.2)(3,0)
	      }
	      \rput[r](4,1.5){$i$}
	      \rput[l](5,1.5){$j$}
	      \rput[l](4.6,0.9){$k$}
	      \rput[l](4.55,1.4){$l$}
	     \end{pspicture}\\
  Pyrochlore
 \end{center}
 \caption{Examples of lattice structure where generalized Hubbard models
 with exact plaquette-ordered ground states can be constructed: the
 one-dimensional chain, the Kagom\'e, the Checkerboard and the
 Pyrochlore lattices. The solid and the dashed plaquettes denote those
 belong to the groups ${\cal A}$ and ${\cal B}$,
 respectively.}\label{fig:lattices}
\end{figure}
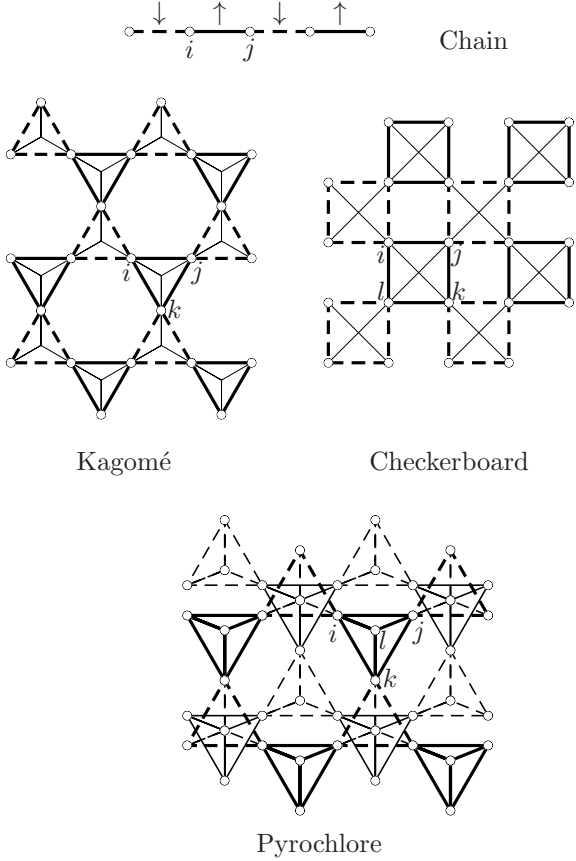

 The above argument is satisfied in corner sharing lattices with the
 bipartite structure in terms of their dual lattices. For example, in
 two dimensions, the Kagom\'e (Checkerboard) lattice can be covered by
 two plaquettes for different sectors alternatively, as illustrated in
 fig.~\ref{fig:lattices}.  These states can be regarded as the N\'eel
 ordering on the honeycomb (square) lattice. In three dimensions, the
 Pyrochlore lattice and the Garnet lattice\cite{Petrenko-P} satisfy
 these conditions.

\section{One-dimensional chain}\label{sec:1D}

First, we consider the one-dimensional generalized Hubbard model at
half-filling and zero-magnetic field, given by ${\cal
 H}=\sum_{i\sigma}h_{i,i+1,\sigma}$ with the local bond Hamiltonian,
\begin{eqnarray}
 \lefteqn{h_{ij\sigma}=-t\,T_{ij\sigma}
  +\frac{U}{2z}
  (n_{i\sigma}n_{i\bar{\sigma}}+n_{j\sigma}n_{j\bar{\sigma}})
  }\nonumber\\
 &&
  +V_{\parallel}n_{i\sigma}n_{j\sigma}+V_{\perp}n_{i\sigma}n_{j\bar{\sigma}}
  \nonumber\\
 &&
  +XT_{ij\sigma}(n_{i\bar{\sigma}}+n_{j\bar{\sigma}})
  +\frac{W}{2}\sum_{\sigma'}T_{ij\sigma}T_{ij{\sigma}'},
  \label{local_bond_Ham}
\end{eqnarray}
where $\bar{\sigma}$ is the opposite spin of $\sigma$, and periodic
boundary conditions are assumed. The number of the nearest sites is
$z=2$.  We have defined the hopping and the density operators as
$T_{ij\sigma}\equiv c_{i\sigma}^{\dag}c_{j\sigma}^{}+\mbox{H.c.}$,
$n_{i\sigma}\equiv c_{i\sigma}^{\dag}c_{i\sigma}^{}$.
Note that the bond-bond interaction ($W$) term can be rewritten as
\begin{equation}
 -2W(\bm{S}_i\cdot\bm{S}_{j}+\bm{\eta}_i\cdot\bm{\eta}_{j}
  -{\textstyle\frac{1}{4}}),\label{eqn:W-term}
\end{equation}
where $\bm{S}_i$ and $\bm{\eta}_i$ are the spin and the pseudo spin
operators, respectively.  The components of the pseudo spin operator are
defined by
\begin{equation}
  \eta_i^{+}\equiv(-1)^i c_{i\uparrow}^{\dag}c_{i\downarrow}^{\dag},\ \
  \eta_i^{-}\equiv(-1)^i c_{i\downarrow}c_{i\uparrow},\ \
  \eta_i^{z}\equiv\frac{1}{2}(n_{i\uparrow}+n_{i\downarrow}-1).
  \label{eqn:eta-pairing}
\end{equation}

\begin{table}[b]
\begin{center}
\begin{tabular}{|c|cccc|}
\hline
 & $|0\rangle$
 & $A_{\sigma}^{\dag}|0\rangle$
 & $B_{\sigma}^{\dag}|0\rangle$
 & $B_{\sigma}^{\dag}A_{\sigma}^{\dag}|0\rangle$\\
\hline
 $1-n_{A\sigma}$  & $1$ & $0$ & $1$ & $0$\\
 $n_{B\sigma}$  & $0$ & $0$ & $1$ & $1$\\
\hline
\end{tabular}
\end{center}
\caption{Projection operators to construct the Hamiltonian on the
 one-dimensional chain with the bond N\'eel (BN) ground state. We can
 also discuss the parameter region of the ferromagnetic (FM) and
 phase-separated (PS) states.
 }  \label{tab:projectors_1D}
\end{table}

Now, we introduce bond operators corresponding to the bonding and the
anti-bonding modes,
\begin{equation}
 A_{ij\sigma}^{\dag}
  ={\textstyle\frac{1}{\sqrt{2}}}
  (c_{i\sigma}^{\dag}+c_{j\sigma}^{\dag}),\quad
 B_{ij\sigma}^{\dag}
  ={\textstyle\frac{1}{\sqrt{2}}}
  (c_{i\sigma}^{\dag}-c_{j\sigma}^{\dag}).
\end{equation}
The two electron states are described as
$B_{ij\sigma}^{\dag}A_{ij\sigma}^{\dag}
 =c_{i\sigma}^{\dag}c_{j\sigma}^{\dag}$.
These operators on the same bond satisfy the anticommutation relations:
\begin{displaymath}
 \{A_{ij\sigma},A_{ij\sigma'}^{\dag}\}
 =\{B_{ij\sigma},B_{ij\sigma'}^{\dag}\}=\delta_{\sigma\sigma'},\quad
 \mbox{otherwise}=0.
\end{displaymath}
The density operators for the bond operators are given as
\begin{eqnarray}
 n_{A\sigma}&\equiv&A_{ij\sigma}^{\dag}A_{ij\sigma}
  =\frac{1}{2}(n_{i\sigma}+n_{j\sigma}+T_{ij\sigma}),\\
 n_{B\sigma}&\equiv&B_{ij\sigma}^{\dag}B_{ij\sigma}
  =\frac{1}{2}(n_{i\sigma}+n_{j\sigma}-T_{ij\sigma}).
\end{eqnarray}
Since we restrict our attention only on the neighboring two sites $i,j$,
we drop these indices from the operators defined above.

\begin{figure}[t]
\begin{center}
 \psset{unit=10mm} 
	    \begin{pspicture}(-4,-0.3)(4,4.5)
	     \pspolygon*[linecolor=lightgray](1.5,3.5)(-1,1)(0.5,0.5)(3,3)
	     \pspolygon*[linecolor=lightgray](3,3)(0.5,0.5)(3,0.5)
	     \pspolygon*[linecolor=lightgray](-3,3)(-0.5,0.5)(-3,0.5)
	     \psgrid[subgriddiv=1,griddots=5,gridlabels=7pt](-4,0)(4,4)
	     \psaxes[labels=none,ticks=none]{->}(0,0)(-4,0)(4,4)
	      \rput[r]{0}(-0.1,3.8){$W/t$}
	      \rput[r]{0}(4,-0.5){$U/2t$}
	      \psline{-}(-1,1)(0.5,0.5)
	      \psline{-}(-1,1)(1.5,3.5)
	      \psline{-}(0.5,0.5)(3,3)
	      \psline{-}(0.5,0.5)(3,0.5)
	      \psline{-}(-3,3)(-0.5,0.5)(-3,0.5)
	      \psline[linestyle=dashed]{-}(0.5,0.5)(-0.25,-0.25)
	      \psline[linestyle=dashed]{-}(-3,1)(3,1)
	      \rput[c]{0}(1.5,3.7){$\lambda_{\bar{A}\bar{A}}=0$}
	      \rput[c]{0}(3.1,3.3){$\lambda_{\bar{A}B}=0$}
	      \rput[l]{0}(-1,-0.5){$\lambda_{BB}=0$}
	     \psline[linewidth=0.5pt,linestyle=solid]{->}(-0.65,-0.3)(-0.1,0.7)
	      \rput[l]{0}(-3.9,3.3){$\lambda_{\bar{A}\bar{A}}+\lambda_{BB}=0$}
	     \rput[c]{0}(1,2){\Large BN}
	     \rput[l]{0}(2,1.5){\Large FM}
	     \rput[r]{0}(-2.1,1.5){\Large PS}
	    \end{pspicture}
 \caption{Phase diagrams of the generalized Hubbard chain
 (\ref{local_bond_Ham}) in the $U/2t$-$W/t$ parameter space with
 $t>0$.\cite{Itoh-N-M,Nakamura-O-I} The other parameters are set as
 $X=t-W$, $V_{\parallel}=W$ and $V_{\perp}=U/2$. The shaded regions
 labelled by BN, FM and PS denote bond-N\'eel, ferromagnetic and
 phase-separated states, respectively.}  \label{phase_diagrams_1D}
\end{center}
\end{figure}
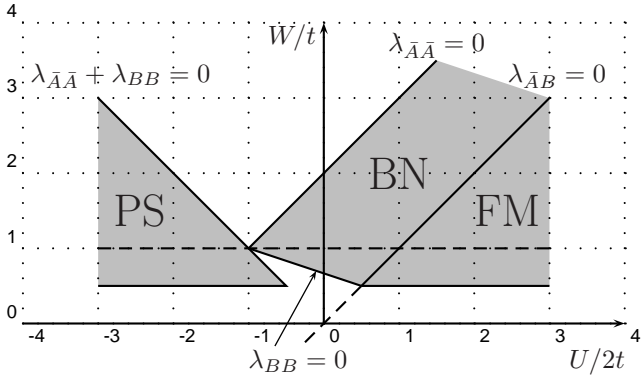

As a trial state, we consider the following wave function,
\begin{equation}
  |\Psi_{\sigma}\rangle\equiv
  A_{12\sigma}^{\dag}
  A_{23\bar{\sigma}}^{\dag}\cdots
  A_{L-1,L\sigma}^{\dag}
  A_{L,1\bar{\sigma}}^{\dag}
  |0\rangle,
\end{equation}
where $|0\rangle$ denotes a vacuum and $L$ is the number of sites.  This
state is regarded as a N\'eel ordering of the bond-located spins, so
that we call this bond N\'eel (BN) state. There is two-fold degeneracy
given by $|\Psi_{\uparrow}\rangle$ and $|\Psi_{\downarrow}\rangle$. In
order to construct a model with the exact ground state, the local
Hamiltonian $h_{ij}=\sum_{\sigma}h_{ij\sigma}$ should be decomposed by
the projection operators $1-n_{A\sigma}$ and $n_{B\sigma}$ (see table
\ref{tab:projectors_1D}) in the following form,
 \begin{eqnarray}
  \lefteqn{h_{ij}-\varepsilon_0
   =\lambda_{\bar{A}\bar{A}}(1-n_{A\uparrow})(1-n_{A\downarrow})
   +\lambda_{BB}n_{B\uparrow}n_{B\downarrow}}\nonumber\\
  &&
   +\lambda_{\bar{A}B}\{(1-n_{A\uparrow})n_{B\downarrow}
   +n_{B\uparrow}(1-n_{A\downarrow})\},
   \label{decomposed_Hamiltonian_1D}
 \end{eqnarray}
where $\varepsilon_0$ is the ground-state energy par bond. According to
the argument given in sec.~\ref{sec:method}, for the BN ground state,
the parameters should be chosen as
\begin{equation}
 \lambda_{\bar{A}\bar{A}},\quad \lambda_{\bar{A}B},\quad\lambda_{BB}
  \geq 0.
\label{condition_of_coefficients}
\end{equation}
Comparing eqs.~(\ref{local_bond_Ham}) and
(\ref{decomposed_Hamiltonian_1D}), the relations among the parameters
are obtained as
\begin{equation}
 V_{\perp}=\frac{U}{2},\quad V_{\parallel}=W,\quad X=t-W.
\end{equation}
The parameters in eq.~(\ref{decomposed_Hamiltonian_1D}) are identified
as follows,
\begin{eqnarray}
 \lambda_{\bar{A}\bar{A}}
  &=&\frac{U}{2}-W+2t,\label{l_AA1}\\
 \lambda_{\bar{A}B}
  &=&-\frac{U}{2}+W,\label{l_AB1}\\
 \lambda_{BB}
  &=&\frac{U}{2}+3W-2t.\label{l_BB1}
\end{eqnarray}
From eqs.~(\ref{condition_of_coefficients}), (\ref{l_AA1}),
(\ref{l_AB1}) and (\ref{l_BB1}), we obtain the parameter space of the BN
ground state as shown in fig.~\ref{phase_diagrams_1D}.  Note that the BN
state appears only for $t>0$ region.

The property of the BN state can be investigated based on the
matrix-product method. According to ref.~\ref{Itoh-N-M}, both
charge-charge and spin-spin correlation functions vanish except for
those of the nearest sites which indicates the existence of the charge
and the spin gaps. On the other hand, the bond-located spin correlation
exhibits a long range order. We can also calculate elementally
excitation spectrum using the matrix-product method as a variational
approach.\cite{Nakamura-O-I}

In the present one-dimensional model at half-filling, we can discuss not
only the BN state but also the ferromagnetic (FM) and the
phase-separated (PS) states. As shown in table \ref{tab:projectors_1D},
the last term of eq.~(\ref{decomposed_Hamiltonian_1D}) stabilizes the
fully polarized FM state for $\lambda_{\bar{A}B}<0$.  Similarly, the PS
state where the system is separated into a domain of doubly occupied
sites and a vacuum, is stabilized when
$\lambda_{\bar{A}\bar{A}}+\lambda_{BB}<0$, neglecting the surface
energy.  As shown in fig.~\ref{phase_diagrams_1D}, the FM and the PS
states appear in the $U/2t$-$W/t$ parameter space symmetrically in the
positive- and in the negative-$U$ regions, respectively. This is
consistent with the fact that the $W$ term is the ferromagnetic exchange
interactions of the spins and the pseudo spins (\ref{eqn:W-term}), and
the PS state is regarded as the FM state of the pseudo-spin space.  The
condition $W/t\geq 1/2$ for the FM and the PS phases is not clearly
obtained in the present argument. This will be discussed in
Sec.~\ref{sec:summary}.

The phase boundary of the BN and the FM states $\lambda_{\bar{A}B}=0$
corresponds to the SU(2) symmetry in the spins $V_{\parallel}=V_{\bot}$,
so that the ground state is highly degenerate.  The system undergoes a
first-order phase transition at this level-crossing point. When $W/t=1$
($X=0$), there is the particle-hole symmetry. At $(U/2t,W/t)=(-1,1)$,
the system has the SU(2) symmetry in the pseudo-spin space, so that the
BN, the PS and the $\eta$-paring states are degenerate.  The other lines
which separate shaded and non-shaded regions in
fig.~\ref{phase_diagrams_1D} do not necessarily mean phase boundaries.

\section{Kagom\'e lattice}\label{sec:2D}

We consider the generalized Hubbard model including three site terms on
the Kagom\'e lattice at $1/3$-filling with zero-magnetic field, ${\cal
H}=\sum_{\langle ijk\rangle\sigma}h_{ijk\sigma}$, where the summation
$\langle ijk\rangle$ is taken in each unit trimer as shown in
fig.~\ref{fig:lattices},
\begin{eqnarray}
 \lefteqn{h_{ijk\sigma}=h_{ij\sigma}+h_{jk\sigma}+h_{ki\sigma}}
  \nonumber\\
 &&
  +W'(T_{ij\sigma}T_{jk\bar{\sigma}}+T_{jk\sigma}T_{ki\bar{\sigma}}
  +T_{ki\sigma}T_{ij\bar{\sigma}})\nonumber\\
 &&
 +X'(T_{ij\sigma}n_{k\bar{\sigma}}+T_{jk\sigma}n_{i\bar{\sigma}}
 +T_{ki\sigma}n_{j\bar{\sigma}}),
 \label{local_trim_Ham}
\end{eqnarray}
where $h_{ij\sigma}$ is the local bond Hamiltonian defined in
eq.~(\ref{local_bond_Ham}) for $z=4$.  The three site interactions
($W'$, $X'$) have the similar structure to those of the $t$-$J$ model.

We define the following one-electron plaquette operators corresponding
to a zero momentum mode, left and right chirality modes,
\begin{eqnarray}
 A_{ijk\sigma}^{\dag}
  &=&{\textstyle\frac{1}{\sqrt{3}}}
  (c_{i\sigma}^{\dag}+c_{j\sigma}^{\dag}+c_{k\sigma}^{\dag}),\\
 B_{ijk\sigma}^{\dag}
  &=&{\textstyle\frac{1}{\sqrt{3}}}
  (c_{i\sigma}^{\dag}+\omega c_{j\sigma}^{\dag}
  +\omega^2 c_{k\sigma}^{\dag}),\\
 C_{ijk\sigma}^{\dag}
  &=&{\textstyle\frac{1}{\sqrt{3}}}
  (c_{i\sigma}^{\dag}+\omega^2c_{j\sigma}^{\dag}
  +\omega c_{k\sigma}^{\dag}),
\end{eqnarray}
where $\omega=\e^{\i 2\pi/3}$.  These operators on the same trimer
satisfy the anticommutation relations:
\begin{displaymath}
 \{A_{ijk\sigma},A_{ijk\sigma'}^{\dag}\}
 =\{B_{ijk\sigma},B_{ijk\sigma'}^{\dag}\}
 =\{C_{ijk\sigma},C_{ijk\sigma'}^{\dag}\}=\delta_{\sigma\sigma'},
\end{displaymath}
and $\mbox{otherwise}=0$. Note that $A_{ijk\sigma}^{\dag}|0\rangle$,
$B_{ijk\sigma}^{\dag}|0\rangle$ and $C_{ijk\sigma}^{\dag}|0\rangle$ are
chosen as eigenstates of density, hopping and current operators:
\begin{eqnarray}
 N_{ijk\sigma}&\equiv&n_{i\sigma}+n_{j\sigma}+n_{k\sigma},\\
 T_{ijk\sigma}&\equiv&T_{ij\sigma}+T_{jk\sigma}+T_{ki\sigma},\\
 J_{ijk\sigma}&\equiv&J_{ij\sigma}+J_{jk\sigma}+J_{ki\sigma},
\end{eqnarray}
where
$J_{ij\sigma}\equiv\i(c_{i\sigma}^{\dag}c_{j\sigma}^{}-\mbox{H.c.})$.
The density operators for the plaquette operators are
\begin{eqnarray}
 n_{A\sigma}&=&\frac{1}{3}
  \left(N_{ijk\sigma}+T_{ijk\sigma}\right),\\
 n_{B\sigma}&=&\frac{1}{6}
 (2N_{ijk\sigma}-T_{ijk\sigma}-\sqrt{3}J_{ijk\sigma}),\\
 n_{C\sigma}&=&\frac{1}{6}
 (2N_{ijk\sigma}-T_{ijk\sigma}+\sqrt{3}J_{ijk\sigma}).
\end{eqnarray}

\begin{figure}[t]
 \psset{unit=10mm}
\begin{center} 
 \begin{pspicture}(-4,-1)(4,3)
  \pspolygon*[linecolor=lightgray]
  (-1.3333333,0.6666666)(0.3333333,0.3333333)(3.0,1.0)(3,1.75)
  \psgrid[subgriddiv=1,griddots=5,gridlabels=7pt](-4,-1)(4,3)
  \psaxes[labels=none,ticks=none]{->}(0,0)(-4,-1)(4,3)
  \rput[r]{0}(-0.1,2.8){$W/t$}  
  \rput[r]{0}(4,-0.3){{$U/2t$}}
  \psline{-}(3.0,1.75)(-1.3333333,0.6666666)(0.3333333,0.3333333)(3.0,1.0)
  \rput[r]{0}(2.2,1.8){$\lambda_{\bar{A}\bar{A}}=0$}
  \rput[c]{0}(2.7,0.3){$\lambda_{\bar{A}B}=0$}
  \rput[r]{0}(-0.6,0.3){$\lambda_{BB}=0$}
  \rput[c]{0}(1.0,0.85){\Large PN}
 \end{pspicture}
 \caption{
 Phase diagram of the generalized Hubbard model on the Kagom\'e
 lattice (\ref{local_trim_Ham}), in the $U/2t$-$W/t$ parameter space
 with $V_{\parallel}=W=W'$, $V_{\perp}=U/2$, $X=X'=t-2W$ and $t>0$.
 The shaded region labelled by PN denotes the plaquette N\'eel state.
 }\label{phase_diagrams_2D}
\end{center}
\end{figure}
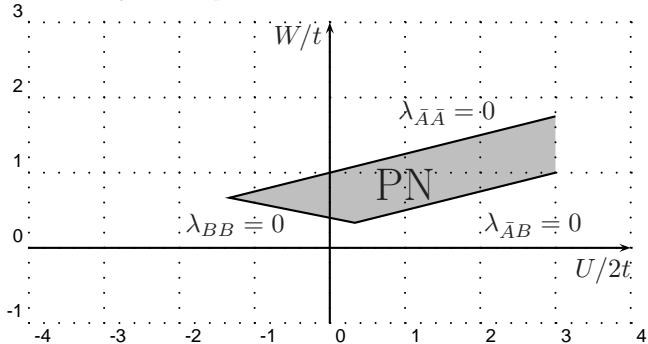

Now we consider the N\'eel ordered state given by $A_{ijk\sigma}^{\dag}$
on the honeycomb lattice as shown in fig.~\ref{fig:lattices}.  Then the
local Hamiltonian $h_{ijk}=\sum_{\sigma}h_{ijk\sigma}$ corresponding to
this state consists of projection operators $1-n_{A\sigma}$,
$n_{B\sigma}$ and $n_{C\sigma}$. In order to eliminate the current term
from the Hamiltonian, we construct $h_{ijk}$ in the following way,
\begin{eqnarray}
 \lefteqn{h_{ijk}-\varepsilon_0=
  \lambda_{\bar{A}\bar{A}}(1- n_{A\uparrow})(1- n_{A\downarrow})}
 \nonumber\\
 &&
  +\lambda_{\bar{A}B}\left\{
  (1-n_{A\uparrow})(n_{B\downarrow}+n_{C\downarrow})
  +(n_{B\uparrow}+n_{C\uparrow})(1-n_{A\downarrow})\right\}
  \nonumber\\
 &&
  +\lambda_{BB}
  (n_{B\uparrow}+n_{C\uparrow})(n_{B\downarrow}+n_{C\downarrow}),
  \label{decomposed_Hamiltonian_2D}
\end{eqnarray}
where the parameters should satisfy the condition of
eq.~(\ref{condition_of_coefficients}). Comparing
eqs.~(\ref{local_trim_Ham}) and (\ref{decomposed_Hamiltonian_2D}), the
relations among the parameters are obtained as
\begin{displaymath}
 V_{\perp}=\frac{U}{2},\quad V_{\parallel}=W=W',\quad 
 X=X'=t-2W,
\end{displaymath}
and parameters in eq.~(\ref{decomposed_Hamiltonian_2D}) are identified
as
\begin{eqnarray}
 \lambda_{\bar{A}\bar{A}}&=&\frac{U}{2}-4W+4t,\\
 \lambda_{\bar{A}B}&=&-\frac{U}{2}+4W-t,\\
 \lambda_{BB}&=&\frac{U}{2}+5W-2t.
\end{eqnarray}
Thus the parameter region of the exact ground state with the plaquette
long-range order is obtained as shown in fig.~\ref{phase_diagrams_2D}.

We can extend the parameter space of the same exact ground state,
introducing more parameters in eq.~(\ref{decomposed_Hamiltonian_2D}),
such as
$\lambda_{BB}(n_{B\uparrow}+n_{C\uparrow})(n_{B\downarrow}+n_{C\downarrow})
\to \lambda_{BB}
(n_{B\uparrow}n_{B\downarrow}+n_{C\uparrow}n_{C\downarrow})
+\lambda_{BB}'
(n_{B\uparrow}n_{C\downarrow}+n_{C\uparrow}n_{B\downarrow})$.  Moreover,
it is also possible to construct models for chiral states given by
$B_{ijk\sigma}^{\dag}$ and $C_{ijk\sigma}^{\dag}$, for the ferromagnetic
state at half-filling, and for two-electron plaquette states at
$2/3$-filling such as
\begin{equation}
 C_{ijk\sigma}^{\dag}B_{ijk\sigma}^{\dag}
  ={\textstyle\frac{\i}{\sqrt{3}}}
  (c_{i\sigma}^{\dag}c_{j\sigma}^{\dag}
  +c_{j\sigma}^{\dag}c_{k\sigma}^{\dag}
  +c_{k\sigma}^{\dag}c_{i\sigma}^{\dag}).
\end{equation}

\begin{table}[b]
\begin{center}
\begin{tabular}{|c|cccc|}
\hline
 & $|0\rangle$
 & $A_{\sigma}^{\dag}|0\rangle$
 & $B_{\sigma}^{\dag}|0\rangle$
 & $B_{\sigma}^{\dag}A_{\sigma}^{\dag}|0\rangle$\\
\hline
 $(1-n_{A\sigma})(1-n_{B\sigma})$ & $1$ & $0$ & $0$ & $0$\\
 $(1-n_{A\sigma})n_{B\sigma}$  & $0$ & $0$ & $1$ & $0$\\
 $n_{A\sigma}n_{B\sigma}$ & $0$ & $0$ & $0$ & $1$\\
\hline
\end{tabular}
 \end{center}
 \caption{
 A different set of projection operators for the
 one-dimensional chain with the BN ground state.}
 \label{tab:projectors_1D-2}
\end{table}

\section{Summary and discussion}\label{sec:summary}

We have discussed a method to construct generalized Hubbard Hamiltonians
with exact plaquette-ordered ground states in arbitrary dimensions. The
corresponding lattices have bipartite structure in terms of corner
sharing unit plaquettes. We have applied this method to the
one-dimensional chain and the Kagom\'e lattice, and obtained parameter
regions of the exact ground states.

In one dimension, the BN state corresponds to the staggered dimer states
in the spin-$1/2$ two-leg ladder model with four spin
exchanges.\cite{Kolezhuk-M} Comparing with the way to decompose the
Hamiltonian to the projection operators in ref.~\ref{Kolezhuk-M}, the
present argument is quite simple. This is because the generalized
Hubbard models we have considered do not include hopping terms between
different spin sectors.

Application of this method to other lattices can be done
straightforwardly.  The bases of plaquette operators should be chosen
reflecting symmetry of lattices. For example, the unit plaquettes of the
Checkerboard and the Pyrochlore lattices consist of four sites, but the
bases of the plaquette operators are chosen in different way.
Similarly, we can also construct Heisenberg-type models with
plaquette-ordered ground states in these lattice systems.  These ground
states have two-fold degeneracy, but the uniqueness has not been proven
yet.

In this paper, we have decomposed particular Hamiltonians into the
projection operators. However, we need more general treatment to
decompose the arbitrary given Hamiltonians.  In the present case, one
projection operator gives positive value for more than two states as
summarized in table.~\ref{tab:projectors_1D}.  To remove this overlap,
we should redefine the projection operators.  For example, in
one-dimensional case, we should choose three operators to treat
arbitrary bond Hamiltonians as shown in table \ref{tab:projectors_1D-2}.
Then the number of free parameters becomes six. On the other hand, in
sec.~\ref{sec:1D}, we have introduced three parameters.  Among these
three extra parameters, the hidden conditions for the ground states
$t>0$ and $W/t\geq 1/2$ are included.  Moreover, when we add the three
site term $T_{ij\sigma}n_{i\bar{\sigma}}n_{j\bar{\sigma}}$ to the model,
these hidden conditions are essential to determine the regions of the BN
ground state.  The detail of this argument will be published
elsewhere\cite{Nakamura-O-I}.

\section*{Acknowledgments}

M.~N. thanks Y.~Motome and K.~Penc for stimulating discussions.
M.~N. is partly supported by the Grant-in-Aid for scientific research of
the Ministry of Education, Science, Sports and Culture of Japan.

\end{document}